\documentclass[conference]{IEEEtran}
\IEEEoverridecommandlockouts
\usepackage{url}
\usepackage{cite}
\usepackage{amsmath,amssymb,amsfonts}
\usepackage{algorithmic}
\usepackage{amsthm}
\usepackage{graphicx}
\usepackage{textcomp}
\usepackage{xcolor}
\usepackage{physics}

\newcommand{\ignore}[1]{}

\newtheorem{lemma}{Lemma}
\newtheorem{prop}{Proposition}
\newtheorem{corollary}{Corollary}
\newtheorem{remark}{Remark}
\newtheorem{definition}{Definition}
\newenvironment{proofsketch}{%
  \proof}{\endproof}
\def\BibTeX{{\rm B\kern-.05em{\sc i\kern-.025em b}\kern-.08em
    T\kern-.1667em\lower.7ex\hbox{E}\kern-.125emX}}
\begin{document}

\title{\bf \LARGE Hybrid Atomic Norm Sparse/Diffuse Channel Estimation \vspace*{-0.1in} \thanks{This work has been funded in part by one or the the  following:  ARO W911NF1910269,
ARO W911NF2410094,
NSF RINGS-2148313,
NSF CCF-2200221,
NSF CCF-2311653,
ONR N00014-22-1-2363, and the
NSF Center for Pandemic Insights DBI-2412522.}}
\author{\IEEEauthorblockN{Lei Lyu and Urbashi Mitra}
\IEEEauthorblockA{\textit{Department of Electrical and Computer Engineering, University of Southern California, CA, USA}}\textit{Email:\{leilyu,ubli\}@usc.edu} \vspace*{-0.1in}}

\maketitle

\begin{abstract}
In this paper, the hybrid sparse/diffuse (HSD) channel model in frequency domain is proposed. Based on the structural analysis on the resolvable paths and diffuse scattering statistics in the channel, the Hybrid Atomic-Least-Squares (HALS) algorithm is designed to estimate sparse/diffuse components with a combined atomic and $l_2$ regularization. A theoretical analysis is conducted on the Lagrangian dual problem and the conditions needed to be satisfied by primal and dual solutions are provided. This analysis, in turn, suggests an algorithm for optimal frequency support estimation. Debiased methods for improved channel estimation are provided. {Given differing amounts of side information, performance bounds are derived in terms of a genie-aided estimator and constrained Cram{\'e}r-Rao lower bounds (CRLB).} Numerical results via simulations on synthetic data as well as real experimental data validate the efficacy of the proposed method.  There are clear tradeoffs with respect to the properties of the channel with respect to performance: sparsity of specular paths and relative energy of diffuse components.
\end{abstract}


\section{Introduction}
Many applications within wireless communications rely on high performance channel estimation. 
For many scenarios, wireless communication channels can be modeled by finite impulse response filters with a sparse number of taps, in these cases, sparse approximation methods yield high performance 
\cite{choi2017compressed,bajwa2010compressed,Tsai,Beygi}
In some environments a mixed model of diffuse and specular, sparse components, as introduced in \cite{michelusi2012uwb,michelusi2012uwb2} is a better fit  (see e.g.\cite{uwbdata,jiang2023long,richter2005joint}).
The HSD model was adapted to underwater acoustic channels in \cite{jiang2023long} and geometry information incorporated into the frequency domain in \cite{santos2010modeling}. Our goal, herein, is to adapt, high performance channel estimation methods based on the atomic norm denoiser \cite{chi2020harnessing,bhaskar2013atomic} to this more realistic channel model.

 
A challenge with the HSD model was that it was proposed for dictionaries which are uniformly sampled (e.g. in time delay or Doppler or angle). Recent work has shown that further improved channel estimation can be achieved if these parameters are unknown, but drawn from a continuous set. To this end, channel estimation methods based on atomic norm denoising \cite{chandrasekaran2012convex,chi2020harnessing,bhaskar2013atomic,LiLOCMAN,Tsai,Beygi}
have proven quite effective.  Herein, we combine the advantages of the atomic norm for estimating sparse, specular components, but introduce a diffuse model to capture the additional components.  We note that direct application of atomic norm methods to the HSD channel does not yield good results, because the differences between the strong specular multipath components and the weaker diffuse components are not taken into consideration.  Thus, in order to provide improvements over a purely atomic norm based model, we uniformly sample the delay space for the diffuse parts.  Our resulting approach is the {\it Hybrid Atomic
Least-Squares algorithm }(HALS).

Deriving a meaningful Cram{\'e}r-Rao bound (CRB) for our signal model is challenging; furthermore, HALS exploits several constraints.  To this end, we adapt prior work on constrained estimators to create two CRBs, which will be used to assess the optimality of HALS.  In particular, we shall consider the constrained CRB for locally $\mathcal{X}$-unbiased estimators derived in \cite{ben2010cramer} for our sparsity constraint, coupled with the norm-constrained CRB \cite{nitzan2019cramer} for our energy constraint.

The main contributions of this paper are: 
\begin{enumerate}
    \item We propose the HSD model \cite{michelusi2012uwb} in continuous time and convert it into a frequency domain model based on the OFDM modulation scheme \cite{hwang2008ofdm}.
    \item Motivated by the demixing algorithm in \cite{mccoy2014convexity}, we propose HALS algorithm that combines the atomic norm \cite{chi2020harnessing,bhaskar2013atomic} and $l_2$ norm to regularize the sparse/diffuse components. The proposed algorithm is validated on both synthetic and experimental data \cite{uwbdata}.  These numerical results show the regimes in which the proposed method offers superior performance to a pure atomic norm approach which presumes a purely sparse channel model.
    \item We derive the dual problems, verify properties that the primal and dual solutions should satisfy.
    {\item Given side information on the subspace of sparse and diffuse components and energy of diffuse components, we derive new constrained CRBs for the proposed problem.} 
\end{enumerate}   

This paper is organized as follows. In Section ~\ref{sec:formulation} we provide the problem formulation.  In particular, Section ~\ref{sec:HSD} describes our signal and channel model.  Section~\ref{sec:ANM} reviews the atomic norm denoising strategy.  Our new algorithm, Hybrid Atomic Least-Square Algorithm (HALS) is introduced in Section~\ref{sec:hals} and connections to the original atomic norm formulation are made showing that the prior formulation is a special case of ours.  Theoretical results, proof sketches and performance bounds to assess the optimality of HALS based on different side information are provided in Sections~\ref{sec:proofs} and~\ref{sec:perform_limit}. 
The algorithm is validated via numerical results in Section~\ref{sec:numerical} based on simulated synthetic channels as well as real channel traces.   The performance of HALS is compared to the derived bounds. Finally, conclusions are drawn in Section~\ref{sec:conclusions}.

\section{Problem Formulation}
\label{sec:formulation}
\subsection{Hybrid Sparse/Diffuse channel Model}
\label{sec:HSD}
We adopt the HSD model in \cite{michelusi2012uwb,michelusi2012uwb2}. Accordingly, the channel impulse response is given by 
\begin{equation}
    h(t) = h_s(t)+h_d(t),
    \label{eq:hsd_model}
\end{equation}
where $h_s(t)$ and $h_d(t)$ represent the sparse and diffuse components in the HSD channel. Denote by $m$ and $L$ the numbers of resolvable multipath components and channel taps, with $m \ll L$. According to \cite{michelusi2011hybrid}, the diffuse components can be regarded as appearing at every tap location. We respectively model the sparse and diffuse components in the channel as
\begin{equation}
    h_s(t)= \sum_{i=1}^m\alpha_i \delta(t-\tau_i); \quad h_d(t)=\sum_{r=0}^{L-1}\gamma_r \delta(t-r\Delta T),
    \label{eq:hsd_model}
\end{equation}
where $\delta\left(\cdot\right)$ is the Dirac delta function, $\alpha_i, \gamma_r \in \mathbb{C}$ are the complex gains of the corresponding paths. $\tau_i \in \mathbb{R}^+$ is the delay time of the i-th resolvable path and $\Delta T \in \mathbb{R}^+$ is the bin interval for the channel taps. 

The OFDM signalling \cite{hwang2008ofdm} is leveraged for pilot signals
herein. To be more specific, the transmitted signal is given by
\begin{equation}
    x(t)=\sum\limits_{k=0}^{N-1} s_k e^{j2\pi k \Delta f t}, \quad 0 \leq t \leq T_s,
    \label{eq:ofdm_transmit}
\end{equation}
where $N$ denotes the number of subcarriers, $s_k \in \mathbb{C}$ is the transmitted symbol over the k-th subcarrier, $\Delta f = \frac{1}{T_s} \in \mathbb{R}$ is the subchannel space and $T_s = L\Delta T \geq \max\left\{\tau_i\right\}$. Assume that cyclic prefix (CP) is employed over $\left[-T_g, 0\right]$ before the transmission, with $T_g$ exceeds the delay spread of the channel.

The received signal through the channel can be written as
\begin{equation}
    y(t) = \sum_{i=1}^m\alpha_i x\left(t-\tau_i\right)+\sum_{r=0}^{L-1}\gamma_r x\left(t-r\Delta T\right) + n(t),
    \label{eq:ofdm_receive}
\end{equation}
where $n(t)$ is the additive noise.

After match filtering and sampling, the discrete signal received over the k-th subcarrier can be expressed as
\begin{align}
    y_k = \frac{1}{T_s}\int_0^{T_s} y(t)e^{-j2\pi k \Delta f t}dt 
    \triangleq h_ks_k+n_k, 
    \label{eq:ofdm_demodulate}
\end{align}

Stacking the signals over subcarriers into vectors, we have
\begin{equation}
    \boldsymbol{y} = \boldsymbol{S}\boldsymbol{h} +\boldsymbol{n},
    \label{eq:ofdm_relation}
\end{equation}
where $\boldsymbol{y}=\left[y_0, y_1, \cdots, y_{N-1}\right]^T$, $\boldsymbol{S} = \text{diag}\left(\boldsymbol{s}\right)$ with $\boldsymbol{s}=\left[s_0, s_1, \cdots, s_{N-1}\right]^T$, $\boldsymbol{h}=\left[h_0, h_1, \cdots, h_{N-1}\right]^T$ and $\boldsymbol{n}=\left[n_0, n_1, \cdots, n_{N-1}\right]^T$.
We next further define the channel model to clearly distinguish the sparse components from the diffuse ones. Define \begin{equation}
    \boldsymbol{a}\left(f\right) \triangleq \left[ 1, e^{j2\pi f}, \cdots, e^{j(N-1)2\pi f} \right]^T.
    \label{eq:atom_nophase}
\end{equation}

The contributions from the sparse components in the HSD channel can be expressed as
\begin{equation}
    \boldsymbol{h}_s = \sum\limits_{i=1}^m \alpha_i \boldsymbol{a}\left(-\frac{\tau_i}{T_s}\right) = \sum\limits_{i=1}^m \alpha_i \boldsymbol{a}\left(\frac{T_s-\tau_i}{T_s}\right) \triangleq \sum\limits_{i=1}^m \alpha_i \boldsymbol{a}\left(f_i\right),
    \label{eq:sparse_channel}
\end{equation}
where $f_i\triangleq \frac{T_s-\tau_i}{T_s} \in \left[0, 1\right]$. As observed in \eqref{eq:sparse_channel}, {assuming $m \ll N$, }$\boldsymbol{h}_s$ will admit sparse decomposition over the set $\left\{\boldsymbol{a}
\left(f\right):f \in[0,1]\right\}$.

Similar to sparse components, diffuse components admit decomposition with respect to \eqref{eq:atom_nophase}
\begin{align}
    \boldsymbol{h}_d& = \sum\limits_{r=0}^{L-1} \gamma_r \boldsymbol{a}\left(-\frac{r\Delta T}{T_s}\right) \nonumber
    \\
    &= \sum\limits_{r=0}^{L-1} \gamma_r \boldsymbol{a}\left(\frac{\left(L-r\right)\Delta T}{L \Delta T}\right) =\sum\limits_{r=0}^{L-1} \gamma_r \boldsymbol{a}\left(\frac{L-r}{L}\right),
    \label{eq:diffuse_channel}
\end{align}
where we assume $T_s = L\Delta T$.

We underscore that the diffuse components here model an aggregate of diffuse components at arbitrary delays.  In general, one could employ a single model with all paths at arbitrary delays; however, to do so would not effectively capture the nature of the channel having both strong specular paths as well as diffuse components.  Furthermore the former approach would yield a channel that is not in fact sparse thus limiting the gains that are typically afforded by the sparse methods.

We thus assume that the diffuse components appear at every channel tap, and construct a basis matrix as follows
\begin{equation}
    \boldsymbol{D} = \frac{1}{\sqrt{N}}\left[\boldsymbol{a}\left(0\right), \boldsymbol{a}\left( \frac{1}{L}\right), \cdots,\boldsymbol{a}\left(\frac{L-1}{L}\right)\right]\in \mathbb{C}^{N\times L},
    \label{eq:diffusebasis}
\end{equation}
with which we can represent the diffuse components by $\boldsymbol{h}_d = \boldsymbol{D}\boldsymbol{c}_d$. Define $\boldsymbol{c}_d \triangleq \sqrt{N}\left[\gamma_0, \gamma_{L-1}, \gamma_{L-2}, \cdots, \gamma_1\right]^T$.
As in \cite{michelusi2012uwb}, we will assume that the diffuse components are of much lower energy  than the specular multipath, thus $\left\Vert\boldsymbol{c}_d\right\Vert_2^2$ will be small. The synthetic channel model and it's dependence on $\beta$ is specified in the Numerical Results section (Section~\ref{sec:numerical}).

\subsection{Atomic norm denoiser}
\label{sec:ANM}
We review the classical atomic norm denoiser in order to set the stage for our proposed method.
The atomic norm, according to \cite{chi2020harnessing}, can be defined as
\begin{align}
    \left\Vert\boldsymbol{x}\right\Vert_\mathcal{A} \triangleq \inf\left\{\sum\limits_{i}c_i:\boldsymbol{x}=c_i\boldsymbol{a}_i, c_i>0, \boldsymbol{a}_i\in \mathcal{A}\right\},
    \label{atomicrewrite}
\end{align}
where $\mathcal{A}$ is some atomic set.

The corresponding dual norm $\left\Vert\boldsymbol{z}\right\Vert^*_\mathcal{A}$ can be defined as
\begin{equation}
    \left\Vert\boldsymbol{z}\right\Vert^*_\mathcal{A} = \sup\limits_{\left\Vert\boldsymbol{x}\right\Vert_{\mathcal{A}}\leq 1} \text{Re}\left<\boldsymbol{z}, \boldsymbol{x}\right>.
\end{equation}

The atomic set we will use is of the form
\begin{align}
    \mathcal{A} &= \left\{\boldsymbol{a}_{f, \phi}=e^{j2\pi \phi}\boldsymbol{a}\left(f\right): f\in\left[0, 1\right], \phi \in \left[0, 1\right]\right\},
    \label{eq:atomicset}
\end{align}
where $\boldsymbol{a}\left(f\right)$ is given in \eqref{eq:atom_nophase}.

Notice that in \eqref{eq:sparse_channel} and 
\eqref{eq:diffuse_channel}, $e^{j2\pi \phi}$, the phase of the atoms, is included in the phase of the complex gains $\alpha_i$ and $\gamma_r$. Particularly, the atomic norm for the sparse components is $\left\Vert\boldsymbol{h}_s\right\Vert_\mathcal{A} = \sum\limits_{i=1}^m \left\vert\alpha_i\right\vert$. The sparsity of solutions can be promoted via atomic norm minimization (ANM) \cite{bhaskar2013atomic,LiLOCMAN}.

\section{Hybrid Atomic-Least-Squares Algorithm}
\label{sec:hals}
Based on the analysis of the structure of sparse and diffuse components, the following optimization problem can be formulated to estimate the HSD channel
\begin{align} \left(\text{P1}\right)\min\limits_{\boldsymbol{h}_s, \boldsymbol{c}_d}&\frac{1}{2}\left\Vert\boldsymbol{S}^{-1}\boldsymbol{y}-\boldsymbol{h}_s-\boldsymbol{D}\boldsymbol{c}_d\right\Vert_2^2 + \tau \left\Vert\boldsymbol{h}_s\right\Vert_{\mathcal{A}}+\frac{\lambda}{2}\left\Vert\boldsymbol{c}_d\right\Vert_2^2,
    \label{eq:als_primal}
\end{align}
where $\tau$ and $\lambda$ are the hyperparameters that need to be determined.

By leveraging the equivalent semidefinite program (SDP) representation of the atomic
norm in \cite{georgiou2007caratheodory} {and the property of the Lagrangian multiplier}, we can reformulate $\left(\text{P1}\right)$ as
\begin{align}
    \left(\text{P2}\right) \min\limits_{t, \boldsymbol{\iota}, \boldsymbol{h}_s, \boldsymbol{c}_d} &\left\Vert\boldsymbol{S}^{-1}\boldsymbol{y}-\boldsymbol{h}_s-\boldsymbol{D}\boldsymbol{c}_d\right\Vert_2^2 + \tau \left(t+\iota_1\right)
    \nonumber
    \\
    \text{s.t.~}  & \left[ \begin{array}{cc} \text{Toep}\left( \boldsymbol{\iota}\right) & \boldsymbol{h}_s \\
\boldsymbol{h}_s^H & t  \end{array}\right] \succeq 0\nonumber
\\
&\left\Vert\boldsymbol{c}_d\right\Vert_2^2 \leq E_d,
\label{eq:als_primal_sdp}
\end{align}
where $\iota_1$ is the first entry of $ \boldsymbol{\iota}$ and $\text{Toep}\left( \boldsymbol{\iota}\right)$ is the Hermitian Toeplitz matrix of the vector $\boldsymbol{\iota}$ {and $E_d$ is the energy estimate on the diffuse components' amplitude}.  We can solve (P2) using off-the-shelf convex solvers, \textit{e.g.} CVX \cite{cvx}, \cite{gb08}. We denote the optimal solution as $\hat{\boldsymbol{h}}_s$ and $\hat{\boldsymbol{c}}_d$. Then, the frequency support estimate of the sparse components, given $\hat{\boldsymbol{h}}_s$ and $\hat{\boldsymbol{c}}_d$, is
\begin{equation}
    \hat{\mathcal{T}} = \left\{\hat{f} \in \left[0, 1\right]:\left\vert\left<\boldsymbol{S}^{-1}\boldsymbol{y}-\hat{\boldsymbol{h}}_s-\boldsymbol{D}\hat{\boldsymbol{c}}_d, \boldsymbol{a}\left(\hat{f}\right)\right>\right\vert = \tau\right\}.
    \label{eq:supportest}
\end{equation}

We construct a basis matrix $\hat{\boldsymbol{G}}$ as $\frac{1}{\sqrt{N}}\left[\boldsymbol{a}\left(\hat{f}\right), \forall \hat{f} \in \hat{\mathcal{T}}\right] \in \mathbb{C}^{N \times \left\vert\hat{\mathcal{T}}\right\vert}$. Then the debiased estimate \cite{bhaskar2013atomic} of the resolvable paths is as
\begin{equation}
    \hat{\boldsymbol{h}}_{s, db} = \hat{\boldsymbol{G}}\hat{\boldsymbol{G}}^\dagger\left(\boldsymbol{S}^{-1}\boldsymbol{y}-\boldsymbol{D}\hat{\boldsymbol{c}}_d\right),
\end{equation}
where $\hat{\boldsymbol{G}}^\dagger$ is the pseudoinverse of $\hat{\boldsymbol{G}}$.  {Note that the debiasing operation is equivalent to projecting the estimate onto the the column space of $\hat{\boldsymbol{G}}$, where we define the projection matrix $\boldsymbol{P}_{\hat{\boldsymbol{G}}}\triangleq\hat{\boldsymbol{G}}\hat{\boldsymbol{G}}^\dagger$, so that $\hat{\boldsymbol{h}}_{s,db}=\boldsymbol{P}_{\hat{\boldsymbol{G}}}\left(\boldsymbol{S}^{-1}\boldsymbol{y}-\boldsymbol{D}\hat{\boldsymbol{c}}_d\right)$. $\boldsymbol{P}_{\hat{\boldsymbol{G}}^\perp}\triangleq\boldsymbol{I}_N-\boldsymbol{P}_{\hat{\boldsymbol{G}}}$ denotes the projection matrix on the corresponding orthogonal complement.}

Our model generalizes the classical purely sparse approach, if we set $\boldsymbol{c}_d = \boldsymbol{0}$ in \eqref{eq:als_primal} and \eqref{eq:als_primal_sdp}, then we recover the ANM methods \cite{bhaskar2013atomic,LiLOCMAN}. We denote this solution as $\hat{\boldsymbol{h}}_{s, ANM}$.

\section{Main Results and Theoretical Analyses}
\label{sec:proofs}
Our key theoretical results assess the correctness of the optimization problem formulation in the context of our new hybrid channel model.
\begin{lemma}[Dual Problem]The dual problem of $\left(\text{P1}\right)$ is
\begin{align}
    \left(\text{P3}\right)\max\limits_{\boldsymbol{z} \in \mathbb{C}^N} & \frac{1}{2}\left\Vert\boldsymbol{S}^{-1}\boldsymbol{y}\right\Vert_2^2 -\frac{1}{2}\left\Vert\boldsymbol{S}^{-1}\boldsymbol{y}-\boldsymbol{z}\right\Vert_2^2-\frac{1}{2\lambda}\left\Vert\boldsymbol{D}^H\boldsymbol{z}\right\Vert_2^2\nonumber
    \\
    \text{s.t.  }& \quad \left\Vert\boldsymbol{z}\right\Vert^*_\mathcal{A} \leq \tau.
    \label{eq:als_dual}
\end{align}

\end{lemma}
\begin{proofsketch} 
Our proof follows along the lines of that for Lemma 2 \cite{bhaskar2013atomic}; however, there are clear differences. In particular,
we need to introduce two constraints $\boldsymbol{x} = \boldsymbol{h}_s$ and $\boldsymbol{b} = \boldsymbol{c}_d$ given the hybrid model. Consequently, we have four primal variables and two dual variables $\boldsymbol{z}$ and $\boldsymbol{v}$ in the Lagrangian $\mathcal{L}\left(\boldsymbol{h}_s, \boldsymbol{c}_d, \boldsymbol{x}, \boldsymbol{b};\boldsymbol{z}, \boldsymbol{v}\right)$. To find the infimum of Lagrangian over primal variables, we rearrange terms into four parts, where the infimum can be taken separately over each part 
\begin{align}
    &\inf\limits_{\boldsymbol{h}_s, \boldsymbol{c}_d, \boldsymbol{x}, \boldsymbol{b}}\mathcal{L}\left(\boldsymbol{h}_s, \boldsymbol{c}_d, \boldsymbol{x}, \boldsymbol{b};\boldsymbol{z}, \boldsymbol{v}\right) \nonumber
    \\
    =&\inf\limits_{\boldsymbol{x}}\underbrace{\left[\tau\left\Vert\boldsymbol{x}\right\Vert_{\mathcal{A}}-\text{Re}\left<\boldsymbol{z}, \boldsymbol{x}\right>\right]}_{\text{(A)}}+\inf\limits_{\boldsymbol{b}}\underbrace{\left[\frac{\lambda}{2}\left\Vert\boldsymbol{b}-\frac{1}{\lambda}\boldsymbol{v}\right\Vert_2^2-\frac{1}{2\lambda}\left\Vert\boldsymbol{v}\right\Vert_2^2\right]}_{\text{(B)}}\nonumber
    \\
    &+\underbrace{\frac{1}{2}\left\Vert\boldsymbol{S}^{-1}\boldsymbol{y}\right\Vert_2^2}_{\text{(C)}}+\inf\limits_{\boldsymbol{h}_s, \boldsymbol{c}_d}\underbrace{\left[\frac{1}{2}\left\Vert\boldsymbol{h}_s+\boldsymbol{D}\boldsymbol{c}_d-\left(\boldsymbol{S}^{-1}\boldsymbol{y}-\boldsymbol{z}\right)\right\Vert_2^2\right.}_{\text{(D)}} \nonumber
    \\
    &\underbrace{\left.-\frac{1}{2}\left\Vert\boldsymbol{S}^{-1}\boldsymbol{y}-\boldsymbol{z}\right\Vert_2^2+\text{Re}\left<\boldsymbol{v}-\boldsymbol{D}^H\boldsymbol{z}, \boldsymbol{c}_d\right>\right]}_{\text{(D)}}.
    \label{eq:als_lag_inf}
\end{align}

Term (C) is constant, whose infimum is itself. Term (A) is a function of $\boldsymbol{x}$ only. According to \cite{bhaskar2013atomic}, the infimum of (A) over $\boldsymbol{x}$ leads to an indicator function, which is equivalent to the constraint $\left\Vert\boldsymbol{z}\right\Vert_\mathcal{A}^*\leq \tau$. We put terms containing $\boldsymbol{b}$ only to (B). To optimize term (B), we complete the squares. The infimum is achieved at $\boldsymbol{b} = \frac{1}{\lambda}\boldsymbol{v}$. All terms containing $\left(\boldsymbol{h}_s, \boldsymbol{c}_d\right)$ are rewritten in part (D). It can be shown that $\boldsymbol{v}-\boldsymbol{D}^H\boldsymbol{z} = \boldsymbol{0}$, or the infimum will go minus infinity. The infimum is thus achieved at $\boldsymbol{h}_s+\boldsymbol{D}\boldsymbol{c}_d = \boldsymbol{S}^{-1}\boldsymbol{y}-\boldsymbol{z}$. Then our dual problem can be formulated as $\left(\text{P3}\right)$.
\end{proofsketch}

\begin{remark}
    Strong duality holds between $\left(\text{P1}\right)$ and $\left(\text{P3}\right)$, since Slater’s condition \cite{boyd2004convex} is satisfied in the constrained $\left(\text{P1}\right)$.
\end{remark}
\begin{lemma}
    The optimal solutions, $\hat{\boldsymbol{h}}_s$ and $\hat{\boldsymbol{c}}_d$, of $\left(\text{P1}\right)$ satisfy:
   \begin{align}
       \left\Vert\boldsymbol{S}^{-1}\boldsymbol{y}-\hat{\boldsymbol{h}}_s-\boldsymbol{D}\hat{\boldsymbol{c}}_d\right\Vert_{\mathcal{A}}^* &\leq \tau
       \label{eq:lemma2_1}
       \\
        \boldsymbol{D}^H\left(\boldsymbol{S}^{-1}\boldsymbol{y}-\hat{\boldsymbol{h}}_s-\boldsymbol{D}\hat{\boldsymbol{c}}_d \right)&= \lambda \hat{\boldsymbol{c}}_d
        \label{eq:lemma2_2}
        \\
        \text{Re}\left<\boldsymbol{S}^{-1}\boldsymbol{y}-\hat{\boldsymbol{h}}_s-\boldsymbol{D}\hat{\boldsymbol{c}}_d, \hat{\boldsymbol{h}}_s\right> &= \tau\left\Vert\hat{\boldsymbol{h}}_s\right\Vert_{\mathcal{A}}.
        \label{eq:lemma2_3}
    \end{align}
\end{lemma}
\begin{proofsketch}
As before, we adapt the proof of Lemma 1\cite{bhaskar2013atomic}, with two key variables to capture our new channel model. Denote the primal objective function as $f\left(\boldsymbol{h}_s, \boldsymbol{c}_d\right)$. Then for any $\boldsymbol{h}_s \in \mathbb{C}^N$, $\boldsymbol{c}_d \in \mathbb{C}^L$ and $\alpha \in \mathbb{R}$, we should have
\begin{equation}
    f\left(\hat{\boldsymbol{h}}_s+\alpha\left(\boldsymbol{h}_s-\hat{\boldsymbol{h}}_s\right), \hat{\boldsymbol{c}}_d+\alpha\left(\boldsymbol{c}_d-\hat{\boldsymbol{c}}_d\right)\right) \geq f\left(\hat{\boldsymbol{h}}_s, \hat{\boldsymbol{c}}_d\right). 
    \label{eq:als_primal_min}
\end{equation}

The convexity of the atomic norm implies \cite{bhaskar2013atomic}
\begin{equation}
    \left\Vert\boldsymbol{h}_s\right\Vert_{\mathcal{A}}-\left\Vert\hat{\boldsymbol{h}}_s\right\Vert_{\mathcal{A}} \geq \alpha^{-1}\left(\left\Vert\hat{\boldsymbol{h}}_s+\alpha\left(\boldsymbol{h}_s-\hat{\boldsymbol{h}}_s\right)\right\Vert_{\mathcal{A}}-\left\Vert\hat{\boldsymbol{h}}_s\right\Vert_{\mathcal{A}}\right).
    \label{eq:atomic_norm_convex}
\end{equation}

With \eqref{eq:als_primal_min} and \eqref{eq:atomic_norm_convex}, we can rearrange the inequality as
\begin{align}
        &\tau\left\Vert\hat{\boldsymbol{h}}_s\right\Vert_{\mathcal{A}}-\text{Re}\left<\boldsymbol{S}^{-1}\boldsymbol{y}-\hat{\boldsymbol{h}}_s-\boldsymbol{D}\hat{\boldsymbol{c}}_d, \hat{\boldsymbol{h}}_s\right>\nonumber
        \\
        \leq &\left(\tau\left\Vert\boldsymbol{h}_s\right\Vert_{\mathcal{A}}-\text{Re}\left<\boldsymbol{S}^{-1}\boldsymbol{y}-\hat{\boldsymbol{h}}_s-\boldsymbol{D}\hat{\boldsymbol{c}}_d, \boldsymbol{h}_s\right>\right)
        \label{eq:lemma2_1_proof}
        \\
        &- \text{Re}\left<\boldsymbol{D}^H\left(\boldsymbol{S}^{-1}\boldsymbol{y}-\hat{\boldsymbol{h}}_s-\boldsymbol{D}\hat{\boldsymbol{c}}_d\right)-\lambda \hat{\boldsymbol{c}}_d, \boldsymbol{c}_d-\hat{\boldsymbol{c}}_d\right>.
        \label{eq:lemma2_2_proof}
\end{align}

{It can be shown that the RHS of the above inequality cannot go minus infinity. If it did, this would contradict the optimality of $\left(\hat{\boldsymbol{h}}_s, \hat{\boldsymbol{c}}_d\right)$.We can leverage this result to conclude that the infimum of the RHS, over $\boldsymbol{h}_s \in \mathbb{C}^N$ and $\boldsymbol{c}_d \in \mathbb{C}^L$, also cannot achieve minus infinity. According to \textbf{Lemma 1}'s proof, the infimum of \eqref{eq:lemma2_1_proof} over $\boldsymbol{h}_s$ is an indicator function that goes to minus infinity unless \eqref{eq:lemma2_1} is true. Similarly, we argue that in \eqref{eq:lemma2_2_proof}, the left entry in the inner product must be $\boldsymbol{0}$, or the infimum over $\boldsymbol{c}_d$ will be minus infinity. This statement proves \eqref{eq:lemma2_2}, which guarantees that the RHS of the above inequality achieves $0$ at $\boldsymbol{h}_s = \boldsymbol{0}$. 
Thus with \eqref{eq:lemma2_1}, the correctness of \eqref{eq:lemma2_3} is verified.}
\end{proofsketch}
\begin{remark}
 $\textbf{Lemma 2}$ motivates the following support estimator: \eqref{eq:supportest}. We can expand the estimate w.r.t. the support estimate $\hat{\mathcal{T}}$ as $\hat{\boldsymbol{h}}_s = \sum\limits_{\hat{f} \in \hat{\mathcal{T}}}c_{\hat{f}}\boldsymbol{a}\left(\hat{f}\right)$. Then \eqref{eq:lemma2_3} can be rewritten as
\begin{eqnarray}
    \sum\limits_{\hat{f} \in \hat{\mathcal{T}}}\text{Re}\left(c_{\hat{f}}^*\left<\boldsymbol{S}^{-1}\boldsymbol{y}-\hat{\boldsymbol{h}}_s-\boldsymbol{D}\hat{\boldsymbol{c}}_d, \boldsymbol{a}\left(\hat{f}\right)\right>\right)=\tau\sum\limits_{\hat{f} \in \hat{\mathcal{T}}}\left\vert c_{\hat{f}}\right\vert,
    \label{eq:inner_prod_exp}
\end{eqnarray}
which, together with \eqref{eq:lemma2_1}, verifies \eqref{eq:supportest}.
\end{remark}

\begin{corollary}
    $\hat{\boldsymbol{z}} = \boldsymbol{S}^{-1}\boldsymbol{y}-\hat{\boldsymbol{h}}_s-\boldsymbol{D}\hat{\boldsymbol{c}}_d$ is a solution to $\left(\text{P3}\right)$.
\end{corollary}
\begin{proofsketch}
    Start from the dual objective function w.r.t. $\hat{\boldsymbol{z}}$. Using $\textbf{Lemma 2}$, we can transfer it into a primal objective function w.r.t. $\hat{\boldsymbol{h}}_s$ and $\hat{\boldsymbol{c}}_d$. Based on the property of the strong duality, $\hat{\boldsymbol{z}}$ is a dual optimal.
\end{proofsketch}
{
\begin{remark}
From \eqref{eq:lemma2_2}, the diffuse component estimator structure is equivalent to a two stage-estimator wherein we are given 
$\hat{\boldsymbol{h}}_s$, and  $\hat{\boldsymbol{c}}_d$ is the optimal solution to the following ridge regression problem:
\begin{equation}
    \min\limits_{\boldsymbol{c}_d \in \mathbb{C}^L} \left\Vert\boldsymbol{S}^{-1}\boldsymbol{y}-\hat{\boldsymbol{h}}_s-\boldsymbol{D}\boldsymbol{c}_d\right\Vert_2^2+\lambda\left\Vert\boldsymbol{c}_d\right\Vert_2^2.
\end{equation}

From \textbf{Corollary 1}, we know that the sparse component estimate from $\left(\text{P1}\right)$ is biased due to the existence of the dual variable. And as to be expected this noisy estimate has an impact on the diffuse component estimator.  This dependence is clearly seen in the ridge regression formulation above.
\end{remark}
}

\begin{remark}
According to \cite{chi2020harnessing}, the dual solution $\hat{\boldsymbol{z}}$ of $\left(\text{P3}\right)$ can be solved by reformulating $\left(\text{P3}\right)$ into an SDP problem.

\end{remark}

{
\begin{remark}
    Denote $\boldsymbol{h}_s^\star$ and $\boldsymbol{D}\boldsymbol{c}_d^\star$ as the true sparse and diffuse components. Given \textbf{Corollary 1}, the estimation error can be equivalently expressed as $\left\Vert\boldsymbol{h}_s^\star+\boldsymbol{D}\boldsymbol{c}_d^\star-\hat{\boldsymbol{h}}_s-\boldsymbol{D}\hat{\boldsymbol{c}}_d\right\Vert_2^2 = \left\Vert\hat{\boldsymbol{z}}-\boldsymbol{w}\right\Vert_2^2$, where $\boldsymbol{w}=\boldsymbol{S}^{-1} \boldsymbol{n}$ is the noise. From the dual problem derived in $\textbf{Lemma 1}$, we see that the magnitude of the dual solution has the following property: $\left\Vert\hat{\boldsymbol{z}}\right\Vert \propto \tau$. Therefore, the solution is more biased, as $\tau$ gets larger. However, according to our experiments,  the impact of debiasing (performance improvement) is also increasing when $\tau$ gets larger. Thus, there is a trade-off between {bias and overall performance}.
\end{remark}
}
{\begin{remark}
    Using \eqref{eq:lemma2_1} and \eqref{eq:lemma2_2} from \textbf{Lemma 2} and \textbf{Corollary 1}, we can derive an upper bound for the estimate of the energy of the diffuse components:
    \begin{align}
        \left\Vert\hat{\boldsymbol{c}}_d\right\Vert_2^2&=\left<\lambda\hat{\boldsymbol{c}}_d, \frac{1}{\lambda}\hat{\boldsymbol{c}}_d\right>= \left<\boldsymbol{D}^H\hat{\boldsymbol{z}}, \frac{1}{\lambda}\hat{\boldsymbol{c}}_d\right>=\frac{1}{\lambda}\left<\hat{\boldsymbol{z}}, \boldsymbol{D}\hat{\boldsymbol{c}}_d\right>\nonumber
        \\
        &\leq \frac{\tau}{\lambda}\left\Vert\boldsymbol{D}\hat{\boldsymbol{c}}_d\right\Vert_\mathcal{A} \leq \frac{\tau}{\lambda}\left\Vert\hat{\boldsymbol{c}}_d\right\Vert_1 \leq \frac{\sqrt{L}\tau}{\lambda}\left\Vert\hat{\boldsymbol{c}}_d\right\Vert_2. \nonumber
    \end{align}
    
Therefore, $\left\Vert\hat{\boldsymbol{c}}_d\right\Vert_2 \leq \frac{\sqrt{L}\tau}{\lambda}$.  This bound is relevant as hyper-parameter tuning remains an interesting open problem for many sparse approximation strategies.  Thus, we expect $\left\Vert\hat{\boldsymbol{c}}_d\right\Vert_2^2 \propto \left\Vert\boldsymbol{c}_d^\star\right\Vert^2_2$, which suggests that we should set $\frac{\sqrt{L}\tau}{\lambda} \propto \left\Vert\boldsymbol{c}_d^\star\right\Vert_2$.

In the synthetic data experiments, we fine-tune $\tau$ around $1.2\sigma \sqrt{N\log{N}}$, which is the suggested hyperparameter value for general ANM problems as derived in \cite{chi2020harnessing}, and $\lambda$ to be approximately $\frac{\sqrt{L}\sigma}{\left\Vert\boldsymbol{c}_d^\star\right\Vert_2}$.  We underscore that derivation of the proper hyperparameter values for HALS is a problem for future work.
\end{remark}
}

{\begin{remark}
    The solution of $\left(\text{P1}\right)$, which is biased, has the error $N \cdot \mbox{MSE}_{b}=\left\Vert\boldsymbol{h}^\star_s+\boldsymbol{D}\boldsymbol{c}^\star_d-\hat{\boldsymbol{h}}_s-\boldsymbol{D}\hat{\boldsymbol{c}}_d\right\Vert_2^2 \triangleq \left\Vert\boldsymbol{e}_{b}\right\Vert_2^2$. Then we can express the debiased error as
\begin{align}
N \cdot \mbox{MSE}_{db} =
&\left\Vert\boldsymbol{h}_s^\star+\boldsymbol{D}\boldsymbol{c}_d^\star-\boldsymbol{P}_{\hat{\boldsymbol{G}}}\left(\boldsymbol{S}^{-1}\boldsymbol{y}-\boldsymbol{D}\hat{\boldsymbol{c}}_d\right)-\boldsymbol{D}\hat{\boldsymbol{c}}_d\right\Vert_2^2\nonumber
    \\
    = & N \cdot \mbox{MSE}_{b}-\left\Vert\boldsymbol{P}_{\hat{\boldsymbol{G}}}\boldsymbol{e}_{b}\right\Vert_2^2+\left\Vert\boldsymbol{P}_{\hat{\boldsymbol{G}}}\boldsymbol{w}\right\Vert_2^2\nonumber
    \\
    = & \left\Vert\boldsymbol{P}_{\hat{\boldsymbol{G}}^\perp}\boldsymbol{h}_s^\star+\boldsymbol{P}_{\hat{\boldsymbol{G}}^\perp}\boldsymbol{D}\left(\boldsymbol{c}_d^\star-\hat{\boldsymbol{c}}_d\right)\right\Vert_2^2+\left\Vert\boldsymbol{P}_{\hat{\boldsymbol{G}}}\boldsymbol{w}\right\Vert_2^2.
    \label{eq:decomp_mse_debiased1}
\end{align}

Ideally, the true sparse-components lie wholly in the subspace of the estimated sparse components.  Interestingly, this expression suggests that the error in the diffuse components should also be in the subspace of the estimated sparse components.  As the dimension of the estimated subspace gets larger,
 the second error term $\left\Vert\boldsymbol{P}_{\hat{\boldsymbol{G}}}\boldsymbol{w}\right\Vert_2^2$ converges to $\left\Vert\boldsymbol{w}\right\Vert^2_2$, which is the error of the simple Least-Squares (LS) solution. Thus we want to operate in a regime where ${\hat{\boldsymbol{h}}_s}$ is more sparse.  A key question is how to further improve the quality of the diffuse components estimate $\hat{\boldsymbol{c}}_d$.
\end{remark}}
\section{Performance Limits}
\label{sec:perform_limit}
Herein, we will consider two analytical bounds to assess the performance of HALS. We note that there are variations on Cram{\'e}r-Rao Bounds that we will consider, based on different assumptions. We begin with key definitions.

\begin{definition}
The equivalent real-valued expression for complex vectors and matrices is given by
\begin{align}
    \tilde{\boldsymbol{\phi}}\triangleq \left[\real\left\{{\boldsymbol{\phi}}\right\}^T, \imaginary\left\{{\boldsymbol{\phi}}\right\}^T\right]^T;
    \tilde{\boldsymbol{\Phi}} \triangleq \left[\begin{array}{cc}
        \real\left\{\boldsymbol{\Phi}\right\} & -\imaginary\left\{\boldsymbol{\Phi}\right\} \\
        \imaginary\left\{\boldsymbol{\Phi}\right\} & \real\left\{\boldsymbol{\Phi}\right\}
    \end{array}\right],
    \label{eq:complex}
\end{align}
where $\boldsymbol{\phi}$ is an arbitrary vector and $\boldsymbol{\Phi}$ is an arbitrary matrix.
\end{definition}
\begin{prop}
    Suppose we receive complex-valued observations $\boldsymbol{\vb{z}}=\boldsymbol{x}+\boldsymbol{\vb{w}}$, where $\boldsymbol{\vb{w}}\sim \mathcal{CN}\left(\boldsymbol{0}, \boldsymbol{K}\right)$ and $\boldsymbol{x}\in \mathbb{C}^N$ is deterministic.
    
It can be rewritten as an equivalent real vector $\tilde{\boldsymbol{\vb{z}}}=\tilde{\boldsymbol{x}}+\tilde{\boldsymbol{\vb{w}}}$,
where $\tilde{\boldsymbol{\vb{w}}}\sim \mathcal{N}\left(\boldsymbol{0}, \frac{1}{2}\tilde{\boldsymbol{K}}\right)$. The fisher information matrix (FIM) of $\tilde{\boldsymbol{x}}$ is $\boldsymbol{J}^{\tilde{\boldsymbol{x}}}=\left(\frac{1}{2}\tilde{\boldsymbol{K}}\right)^{-1}$. We see that
\begin{equation}
    \mathbb{E}\left[\left\Vert\hat{\boldsymbol{x}}-\boldsymbol{x}\right\Vert_2^2\right]=\mathbb{E}\left[\left\Vert\hat{\tilde{\boldsymbol{x}}}-\tilde{\boldsymbol{x}}\right\Vert_2^2\right]\geq \mbox{trace}\left\{\left(\boldsymbol{J}^{\tilde{\boldsymbol{x}}}\right)^{-1}\right\}.\label{eq:real_complex_normrealtion}
\end{equation}
\end{prop}
\begin{definition}
    Given the  constraint set,
    \begin{align}
        \mathcal{X}&=\left\{\boldsymbol{x}\in\mathbb{R}^N:\boldsymbol{f}\left(\boldsymbol{x}\right)=\boldsymbol{0}\right\},
    \end{align}
    and $\boldsymbol{F}\left(\boldsymbol{x}\right)\triangleq \nabla_{\boldsymbol{x}}\boldsymbol{f}\left(\boldsymbol{x}\right) \in \mathbb{R}^{K \times N}$ and assume it is of rank $K$. There exists a matrix $\boldsymbol{U}\left(\boldsymbol{x}\right)\in \mathbb{R}^{N\times\left(N-K\right)}$ satisfying
\begin{align}
\boldsymbol{F}\left(\boldsymbol{x}\right)\boldsymbol{U}\left(\boldsymbol{x}\right) & = \boldsymbol{0}\mbox{;~}\boldsymbol{U}^T\left(\boldsymbol{x}\right)\boldsymbol{U}\left(\boldsymbol{x}\right) = \boldsymbol{I}_{N-K}.
    \label{eq:u_mat_def}
\end{align}
\end{definition}
\begin{remark}
    A complex subspace constraint $\boldsymbol{\theta}\in \mathcal{R}\left(\boldsymbol{\Theta}\right)$, for $\boldsymbol{\Theta}\in \mathbb{C}^{N \times M}$, can be equivalently expressed as
    \begin{equation}
        \tilde{\boldsymbol{\theta}}\in\mathcal{R}\left(\tilde{\boldsymbol{\Theta}}\right)\Leftrightarrow \boldsymbol{N}^T_{\tilde{\boldsymbol{\Theta}}^T}\tilde{\boldsymbol{\theta}}=\boldsymbol{0}\Leftrightarrow\tilde{\boldsymbol{N}}^T_{\boldsymbol{\Theta}^H}\tilde{\boldsymbol{\theta}}=\boldsymbol{0},
        \label{eq:sub_equiv}
    \end{equation}
    where the columns of $\boldsymbol{N}_{\tilde{\boldsymbol{\Theta}}^T}$ and $\boldsymbol{N}_{\boldsymbol{\Theta}^H}$ span the null space of $\tilde{\boldsymbol{\Theta}}^T$ and $\boldsymbol{\Theta}^H$.
\end{remark}
We next provide an approximate bound on the MSE for the sparse components. This bound is approximate due to the fact that we will model the diffuse components as additional noise versus structured unknowns.
\begin{definition}[Locally $\mathcal{X}$-unbiasedness conditions\cite{ben2010cramer}]
An estimator $\hat{\boldsymbol{x}}\left(\boldsymbol{\vb{y}}\right)$ is said to be locally $\mathcal{X}$-unbiased in the vicinity of $\boldsymbol{x}_0 \in \mathcal{X}$, if it satisfies the conditions
\begin{align}
    \boldsymbol{b}_{\hat{\boldsymbol{x}}}\left(\boldsymbol{x}_0\right)&=\boldsymbol{0}\mbox{;~ }\boldsymbol{B}_{\hat{\boldsymbol{x}}}\left(\boldsymbol{x}_0\right)\boldsymbol{U}\left(\boldsymbol{x}_0\right)=\boldsymbol{0},
\end{align}
where we denote
\begin{align}
    \boldsymbol{b}_{\hat{\boldsymbol{x}}}\left(\boldsymbol{x}\right)&\triangleq \mathbb{E}\left[\hat{\boldsymbol{x}}\left(\boldsymbol{\vb{y}}\right)-\boldsymbol{x}\right];\boldsymbol{B}_{\hat{\boldsymbol{x}}}\left(\boldsymbol{x}\right)\triangleq \nabla_{\boldsymbol{x}}\boldsymbol{b}_{\hat{\boldsymbol{x}}}\left(\boldsymbol{x}\right).
\end{align}
\end{definition}
\begin{definition}[Locally C-unbiasedness conditions\cite{nitzan2018cram, nitzan2019cramer}]
    An estimator $\hat{\boldsymbol{x}}\left(\boldsymbol{\vb{y}}\right)$ is said to be locally C-unbiased in the vicinity of $\boldsymbol{x}_0 \in \mathcal{X}$, if it satisfies the conditions
    \begin{align}
        \boldsymbol{U}^T\left(\boldsymbol{x}_0\right)\boldsymbol{b}_{\hat{\boldsymbol{x}}}\left(\boldsymbol{x}_0\right) &= \boldsymbol{0};
        \\
        \boldsymbol{u}_k^T\left(\boldsymbol{x}_0\right)\boldsymbol{B}_{\hat{\boldsymbol{x}}}\left(\boldsymbol{x}_0\right)\boldsymbol{U}\left(\boldsymbol{x}_0\right)&=-\boldsymbol{b}^T_{\hat{\boldsymbol{x}}}\left(\boldsymbol{x}_0\right) \nabla_{\boldsymbol{x}}\boldsymbol{u}_k\left(\boldsymbol{x}_0\right)\boldsymbol{U}\left(\boldsymbol{x}_0\right),
    \end{align}
    where $\boldsymbol{u}_k\left(\boldsymbol{x}_0\right)$ is the k-th column of the matrix $\boldsymbol{U}\left(\boldsymbol{x}_0\right)$.
\end{definition}
\begin{prop}[CRB on $\boldsymbol{h}_s$]
    Let $\hat{\tilde{\boldsymbol{h}}}_s\triangleq\hat{\tilde{\boldsymbol{h}}}_s\left(\vb{\boldsymbol{y}}\right)$ be a $\mathcal{X}$-unbiased estimator on $\tilde{\boldsymbol{h}}_s$, the equivalent real vector of complex parameters $\boldsymbol{h}_s\in \mathcal{R}\left(\boldsymbol{G}\right)$ given the observation model in  \eqref{eq:ofdm_relation}, where $\mathcal{R}\left(\boldsymbol{G}\right)$ denotes the range of sparse components support matrix $\boldsymbol{G}\in \mathbb{C}^{N \times m}$. Furthermore, we assume  $\vb{n}\sim \mathcal{CN}\left(\boldsymbol{0}, \sigma^2 \boldsymbol{I}\right)$ and the amplitude of the diffuse components  are distributed as $\boldsymbol{\vb{c}}_d\sim\mathcal{CN}\left(\boldsymbol{0}, \boldsymbol{K}_d\right)$.    
Given the matrix $\boldsymbol{U} \in \mathbb{R}^{2N \times 2m}$ that satisfies
    \begin{equation}
\mathcal{R}\left(\boldsymbol{U}\right)=\mathcal{R}\left(\tilde{\boldsymbol{G}}\right)\mbox{;~} \boldsymbol{U}^T\boldsymbol{U}=\boldsymbol{I}_{2m},
    \end{equation}
    and assuming $\mathcal{R}\left(\boldsymbol{U}\boldsymbol{U}^T\right)\subseteq\mathcal{R}\left(\boldsymbol{U}\boldsymbol{U}^T\left(\tilde{\boldsymbol{K}}_{s,0}\right)^{-1}\boldsymbol{U}\boldsymbol{U}^T\right)$, then
    \begin{align}
\mathbb{E}\left[\left\Vert\hat{\boldsymbol{h}}_s-\boldsymbol{h}_s\right\Vert_2^2\right] \geq\frac{1}{2}\mbox{trace}\left\{\boldsymbol{U}\left(\boldsymbol{U}^T\left(\tilde{\boldsymbol{K}}_{s,0}\right)^{-1}\boldsymbol{U}\right)^\dagger\boldsymbol{U}^T\right\},
        \label{eq:sparse_crb}
    \end{align}
    where $\boldsymbol{K}_{s, 0}\triangleq\boldsymbol{D}\boldsymbol{K}_d\boldsymbol{D}^H+\sigma^2\boldsymbol{S}^{-1}\left(\boldsymbol{S}^{-1}\right)^H$. The relationship between $\boldsymbol{K}_{s, 0}$  and $\tilde{\boldsymbol{K}}_{s,0}$ is given in Equation ~\eqref{eq:complex}.
\end{prop}
\begin{proofsketch}
We will adapt the FIM expression in \eqref{eq:real_complex_normrealtion}, to incorporate constraints such that we can make a statement about the bound on the MSE for  $\tilde{\boldsymbol{h}}_s \in \mathbb{R}^{2N}$.  We first map the general subspace constraint in \eqref{eq:sub_equiv} to our specific problem statement.  That is, 
$\boldsymbol{h}_s\in \mathcal{R}\left(\boldsymbol{G}\right)$  in  \eqref{eq:sub_equiv},  is equivalent to $\boldsymbol{N}^T_{\tilde{G}^T}\tilde{\boldsymbol{h}}_s = \boldsymbol{0}$. The $\mathbf{F}$ matrix in \eqref{eq:u_mat_def} is $\boldsymbol{N}^T_{\tilde{G}^T}$ in this problem. Thus, we know that $\boldsymbol{U}$ is the orthonormal basis matrix of $\mathcal{R}\left(\tilde{\boldsymbol{G}}\right)$.

Treating the diffuse components as additional noise, the observations $\boldsymbol{S}^{-1}\boldsymbol{\vb{y}} = \boldsymbol{h}_s +\boldsymbol{D}\boldsymbol{\vb{c}}_d+\boldsymbol{S}^{-1}\boldsymbol{\vb{n}}$ can be regarded as $\boldsymbol{h}_s$ plus random Gaussian noise with the covariance matrix $\boldsymbol{K}_{s,0}=\boldsymbol{D}\boldsymbol{K}_d\boldsymbol{D}^H+\sigma^2\boldsymbol{S}^{-1}\left(\boldsymbol{S}^{-1}\right)^H$. According to \textbf{Proposition 1}, the FIM $\boldsymbol{J}^{\tilde{\boldsymbol{h}}_s}=\left(\frac{1}{2}\tilde{\boldsymbol{K}}_{s,0}\right)^{-1}$.

With the FIM and the appropriately defined $\mathbf{U}$ matrix, we can substitute into the formula in \textbf{Theorem 1} \cite{ben2010cramer}. This yields the desired result \eqref{eq:sparse_crb}.
\end{proofsketch}
We now provide an approximate bound on the MSE for the diffuse components, exploiting the same type of approximation as above on the sparse components.
\begin{prop}[CRB on $\boldsymbol{h}_d$]
    Let $\hat{\tilde{\boldsymbol{h}}}_d\triangleq\hat{\tilde{\boldsymbol{h}}}_d\left(\vb{\boldsymbol{y}}\right)$ be a C-unbiased estimator on $\tilde{\boldsymbol{h}}_d$, the equivalent real vector of parameters
    \begin{equation}
        \boldsymbol{h}_d\in \mathcal{H}_d=\left\{\boldsymbol{h}_d\in\mathcal{R}\left(\boldsymbol{D}\right):\left\Vert\boldsymbol{h}_d\right\Vert_2^2=\rho^2\right\}.
        \label{eq:nccon_set}
    \end{equation}
    Assume the noise $\vb{n}\sim \mathcal{CN}\left(\boldsymbol{0}, \sigma^2 \boldsymbol{I}\right)$ and the amplitude of sparse components $\boldsymbol{\vb{c}}_s\sim\mathcal{CN}\left(\boldsymbol{0}, \boldsymbol{K}_s\right)$.
    Given the matrix $\boldsymbol{U}_n \in \mathbb{R}^{2N \times (2L-1)}$ that satisfies
\begin{align}
     \left[\begin{array}{c}
\tilde{\boldsymbol{N}}^T_{\boldsymbol{D}^H} \\
          \tilde{\boldsymbol{h}}_d^T
    \end{array}\right] \boldsymbol{U}_n = \boldsymbol{0}\mbox{; ~} \boldsymbol{U}^T_n\boldsymbol{U}_n = \boldsymbol{I}_{2L-1},
    \label{eq:norm_con_umat}
\end{align}
where the columns of $\boldsymbol{N}_{\boldsymbol{D}^H}$ span the null space of $\boldsymbol{D}^H$.

We further assume the following three conditions are satisfied  (first two from \cite{nitzan2019cramer}):
\begin{enumerate}
    \item Integration with respect to $\boldsymbol{\vb{y}}$ and differentiation with respect to $\boldsymbol{h}_d$ is interchangeable.
    \item $\boldsymbol{U}_n^T\left(\tilde{\boldsymbol{K}}_{d, 0}\right)^{-1}\boldsymbol{U}_n$ is nonzero and of finite size.
\item and
$\forall \boldsymbol{h}_d \in \mathcal{H}_d$, 
\begin{align}
\mathbb{E}\left[\left\Vert\hat{\boldsymbol{h}}_d\right\Vert_2^2\right] =\rho^2\mbox{;~} \hat{\boldsymbol{h}}_d\in \mathcal{R}\left(\boldsymbol{D}\right).\label{eq:nc_est_set}
\end{align}
\end{enumerate}
Then,
\begin{align}
    \mathbb{E}\left[\left\Vert\hat{\boldsymbol{h}}_d-\boldsymbol{h}_d\right\Vert_2^2\right] \geq 2\rho^2\left(1-\frac{1}{\sqrt{1+\epsilon/2\rho^2}}\right)
    \label{eq:diffuse_crb},
\end{align}
where $\epsilon\triangleq \mbox{trace}\left\{\boldsymbol{U}_n\left(\boldsymbol{U}_n^T\left(\tilde{\boldsymbol{K}}_{d,0}\right)^{-1}\boldsymbol{U}_n\right)^\dagger\boldsymbol{U}_n^T\right\}$ and $\boldsymbol{K}_{d, 0}\triangleq\boldsymbol{G}\boldsymbol{K}_s\boldsymbol{G}^H+\sigma^2\boldsymbol{S}^{-1}\left(\boldsymbol{S}^{-1}\right)^H$.
\end{prop}

\begin{proofsketch}
Starting from \eqref{eq:sub_equiv}, the needed constraint matrix $\boldsymbol{U}_n \triangleq\boldsymbol{U}_n\left(\tilde{\boldsymbol{h}}_d\right)$ satisfying \eqref{eq:norm_con_umat} can be shown to have the property
\begin{equation}
    \boldsymbol{U}_n\boldsymbol{U}^T_n = \boldsymbol{P}_{\tilde{\boldsymbol{D}}}-\frac{1}{\rho^2}\tilde{\boldsymbol{h}}_d\tilde{\boldsymbol{h}}_d^T.
    \label{eq:umat_proj}
\end{equation}

Showing the desired result hinges on bounding the following quantities: $\mathbb{E}\left[\hat{\tilde{\boldsymbol{h}}}_d^T\boldsymbol{U}_n \boldsymbol{U}^T_n\hat{\tilde{\boldsymbol{h}}}_d\right]$ and $\mathbb{E}\left[\hat{\tilde{\boldsymbol{h}}}_d^T\left(\frac{1}{\rho^2}\tilde{\boldsymbol{h}}_d\tilde{\boldsymbol{h}}_d^T\right)\hat{\tilde{\boldsymbol{h}}}_d\right]$. 
The Cauchy-Schwarz and Jensen's inequality are applied \cite{nitzan2019cramer} respectively. In fact,
the key difference between the derivations in \cite{nitzan2019cramer} and ours herein, are the steps to integrate our particular constraints, \eqref{eq:nccon_set} and \eqref{eq:nc_est_set}, and the projection operation \eqref{eq:umat_proj} characterized by matrix $\boldsymbol{U}_n$. Specifically, we make changes in the proof when we derive the expression for $\boldsymbol{b}_{\hat{\tilde{\boldsymbol{h}}}_d}\left(\tilde{\boldsymbol{h}}_d\right)$ and combine the bounds computed for those two quantities. 

We derive the lower bound in \eqref{eq:diffuse_crb}, where the FIM $\boldsymbol{J}^{\tilde{\boldsymbol{h}}_d}$ is derived by treating the sparse components as additional noise. The observations $\boldsymbol{S}^{-1}\boldsymbol{\vb{y}} = \boldsymbol{h}_d +\boldsymbol{G}\boldsymbol{\vb{c}}_s+\boldsymbol{S}^{-1}\boldsymbol{\vb{n}}$ can be regarded as $\boldsymbol{h}_d$ plus random Gaussian noise with the covariance matrix $\boldsymbol{K}_{d,0}=\boldsymbol{G}\boldsymbol{K}_s\boldsymbol{G}^H+\sigma^2\boldsymbol{S}^{-1}\left(\boldsymbol{S}^{-1}\right)^H$. According to \textbf{Proposition 1}, the FIM $\boldsymbol{J}^{\tilde{\boldsymbol{h}}_d}=\left(\frac{1}{2}\tilde{\boldsymbol{K}}_{d,0}\right)^{-1}$.
\end{proofsketch}

We shall examine these two approximate CRBs \emph{vis \`a vis} the performance of HALS in the Numerical Results. We finally, describe a genie-aided estimator with significant side information.
\begin{prop}[Genie-aided estimator]
    We construct a  genie-aided scheme which, in addition to knowledge of the support of the diffuse components, $\boldsymbol{D}$, also has knowledge of the support of the sparse components, $\boldsymbol{G}$ and the expected energy of the diffuse components, $E_d$:
\begin{align}
    \left(\text{P4}\right)\min\limits_{\boldsymbol{c}_{ss}, \boldsymbol{c}_{dd}} & \left\Vert\boldsymbol{S}^{-1}\boldsymbol{y}-\boldsymbol{G}\boldsymbol{c}_{ss}-\boldsymbol{D}\boldsymbol{c}_{dd}\right\Vert_2^2+\mu\left\Vert\boldsymbol{c}_{dd}\right\Vert_2^2.
\end{align}

The genie estimator has a closed-form solution on channel parameters, given $\hat{\boldsymbol{h}}_{s, ge} = \boldsymbol{G}\hat{\boldsymbol{c}}_{ss}$ and $\hat{\boldsymbol{h}}_{d, ge} = \boldsymbol{D}\hat{\boldsymbol{c}}_{dd}$
\begin{equation}
  \left[\begin{array}{c}
         \hat{\boldsymbol{h}}_{s, ge}  \\
          \hat{\boldsymbol{h}}_{d, ge}\end{array}\right]= \left[\begin{array}{c}
         \boldsymbol{P}_{\boldsymbol{G}}\left(\boldsymbol{I}-\boldsymbol{D}\boldsymbol{T}^{-1}\boldsymbol{D}^H\boldsymbol{P}_{\boldsymbol{G}^{\perp}}\right)  \\
          
    \boldsymbol{D}\boldsymbol{T}^{-1}\boldsymbol{D}^H\boldsymbol{P}_{\boldsymbol{G}^{\perp}}\end{array}\right]\boldsymbol{S}^{-1}\boldsymbol{y},
    \label{eq:genie_est}
\end{equation}
where $\boldsymbol{P}_{\boldsymbol{G}} = \boldsymbol{G}\boldsymbol{G}^{\dagger}$, $\boldsymbol{P}_{\boldsymbol{G}^{\perp}} = \boldsymbol{I}-\boldsymbol{P}_{\boldsymbol{G}}$ and $\boldsymbol{T}\triangleq \boldsymbol{D}^H\boldsymbol{P}_{\boldsymbol{G}^{\perp}}\boldsymbol{D}+\mu\boldsymbol{I}$.
\end{prop}

The performance of the genie-aided algorithm will be used as a benchmark against our HALS estimator as well as the approximate CRBs derived above in the sequel.

\section{Numerical Results}
\label{sec:numerical}
We first consider purely synthetic channels.
The sampling rate is such that $\Delta T \cdot \Delta f = \frac{1}{L}$. We generate random values $\vb{\tau}_i$ from $\text{Uniform}\left(0, L\Delta T\right)$ independently for $i = 1, \cdots, m$. We assume Rayleigh fading \cite{michelusi2012uwb} for the path gains. The complex channel gain $\alpha_i$ is drawn from $\mathcal{CN}\left(0, e^{-\omega\frac{\tau_i}{\Delta T}}\right)$ for $i = 1, \cdots, m$. The complex gain $\gamma_r$ is generated from $\mathcal{CN}\left(0, \beta e^{-\omega r}\right)$ for $r = 0, \cdots, L-1$.
The noise is drawn from $\boldsymbol{\vb{n}} \sim \mathcal{CN}\left(\boldsymbol{0}, \sigma^2 \boldsymbol{I}_N\right)$.  {Our data symbols are drawn uniformly from the QPSK constellation.}
We also examine the performance of HALS using real channel traces measured in the frequency domain  (2-8 GHz) from the NIST `Child Care'' dataset \cite{uwbdata}. The 2nd and 13th measurements are denoted as CC2 and CC13.  {The CC2 and CC13 channel responses differ in the amount of specular mulitpath and relative magnitudes of the diffuse components with respect to the specular components.} {The CC2 channel has a greater number of sparse components and more significant diffuse scattering. Then, we expect our algorithm to perform much better on estimating the CC13 channel, which is observed in the experiments.}

For the synthetic channels, $\boldsymbol{D}$ is  known.
We define the signal-to-noise ratio, $\mbox{SNR(dB)} = 10\log_{10}\frac{\left\Vert\boldsymbol{h}^\star\right\Vert_2^2}{N\sigma^2}$.  Our performance metric is the normalized mean square error (NMSE): $\mbox{NMSE}=\frac{\left\Vert\boldsymbol{h}^\star-\hat{\boldsymbol{h}}\right\Vert_2^2}{\left\Vert\boldsymbol{h}^\star\right\Vert_2^2}$, where $\boldsymbol{h}^\star$ is the true channel and $\hat{\boldsymbol{h}}$ is the estimated channel. 
For our HALS algorithm, $\hat{\boldsymbol{h}}_{HALS} = \hat{\boldsymbol{h}}_{s, db}+\boldsymbol{D}\hat{\boldsymbol{c}}_d$. The vanilla ANM method is implemented by setting $\boldsymbol{c}_d = \boldsymbol{0}$ in \eqref{eq:als_primal_sdp}, which gives an estimate $\hat{\boldsymbol{h}}_{ANM}=\hat{\boldsymbol{h}}_{s,ANM}$. As a reference, the genie-aided (Genie) scheme in \eqref{eq:genie_est} is implemented, which provides an estimate $\hat{\boldsymbol{h}}_{ge} = \hat{\boldsymbol{h}}_{s,ge}+\hat{\boldsymbol{h}}_{d,ge}$. The hyper-parameters for these methods are all fine-tuned. The least square (LS) solution can be formulated as $\hat{\boldsymbol{h}}_{LS}=\boldsymbol{S}^{-1}\boldsymbol{y}$. The convex problems are solved by CVX \cite{cvx,gb08}. For sparse components and diffuse components, the MSE and CRB results are normalized by dividing $\left\Vert\boldsymbol{h}_s^\star\right\Vert_2^2$ and $\left\Vert\boldsymbol{h}_d^\star\right\Vert_2^2$ respectively. The normalized CRBs (NCRB), which are compared with sparse and diffuse components estimators from the HALS and Genie scheme, are computed for sparse components \eqref{eq:sparse_crb} as well as diffuse components \eqref{eq:diffuse_crb}.

\begin{figure}[t!]
  \centerline{\includegraphics[width=8.5cm]{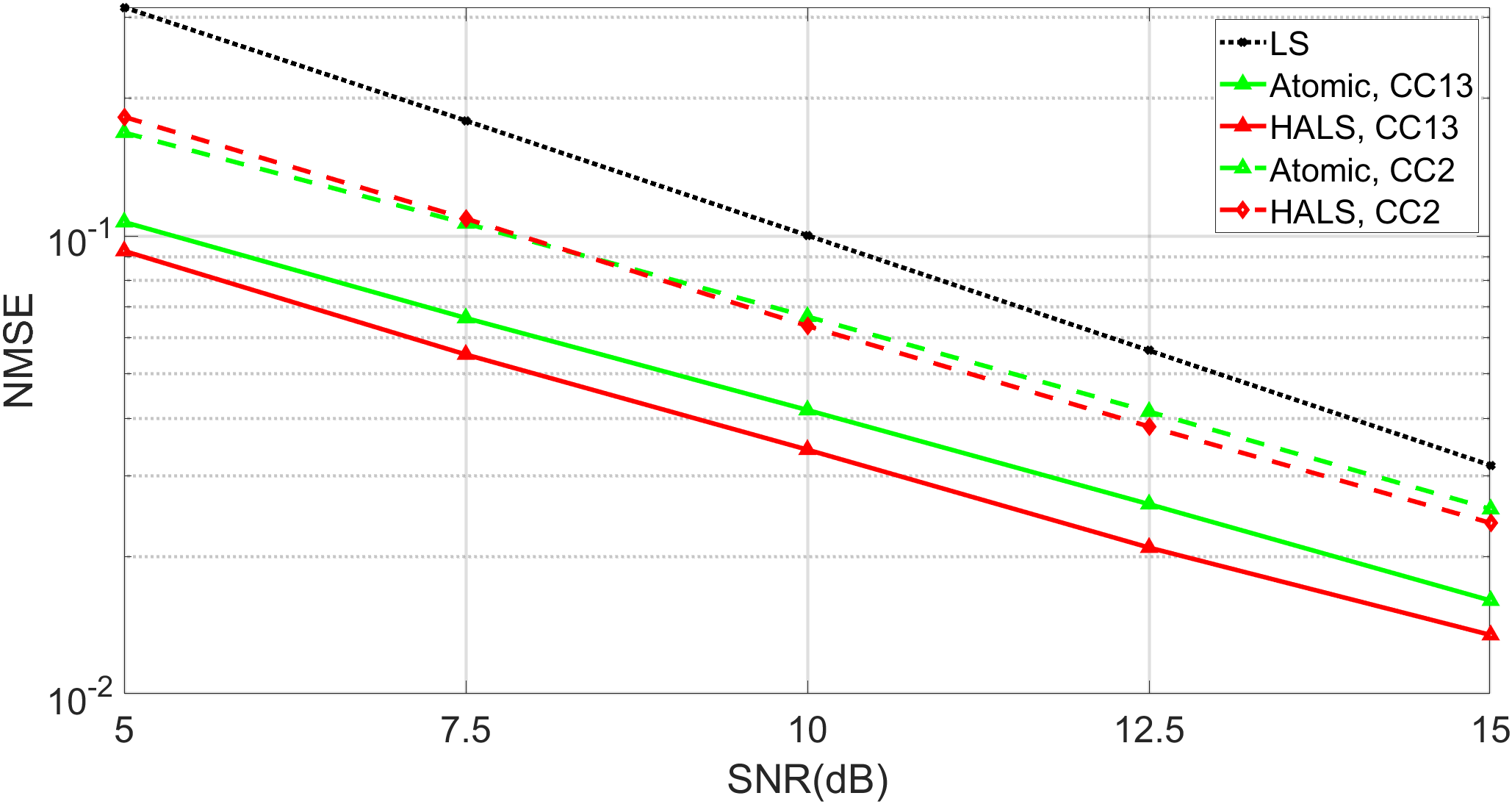}}
    \vspace*{0.1in}
\caption{NMSE versus SNR results with real data. Parameter values are: $N=50$ and 
 $\tilde{L} =200$ for real channel estimation.}
\label{fig:exp1_1}
\end{figure}
\begin{figure}[t!]
  \centerline{\includegraphics[width=8.5cm]{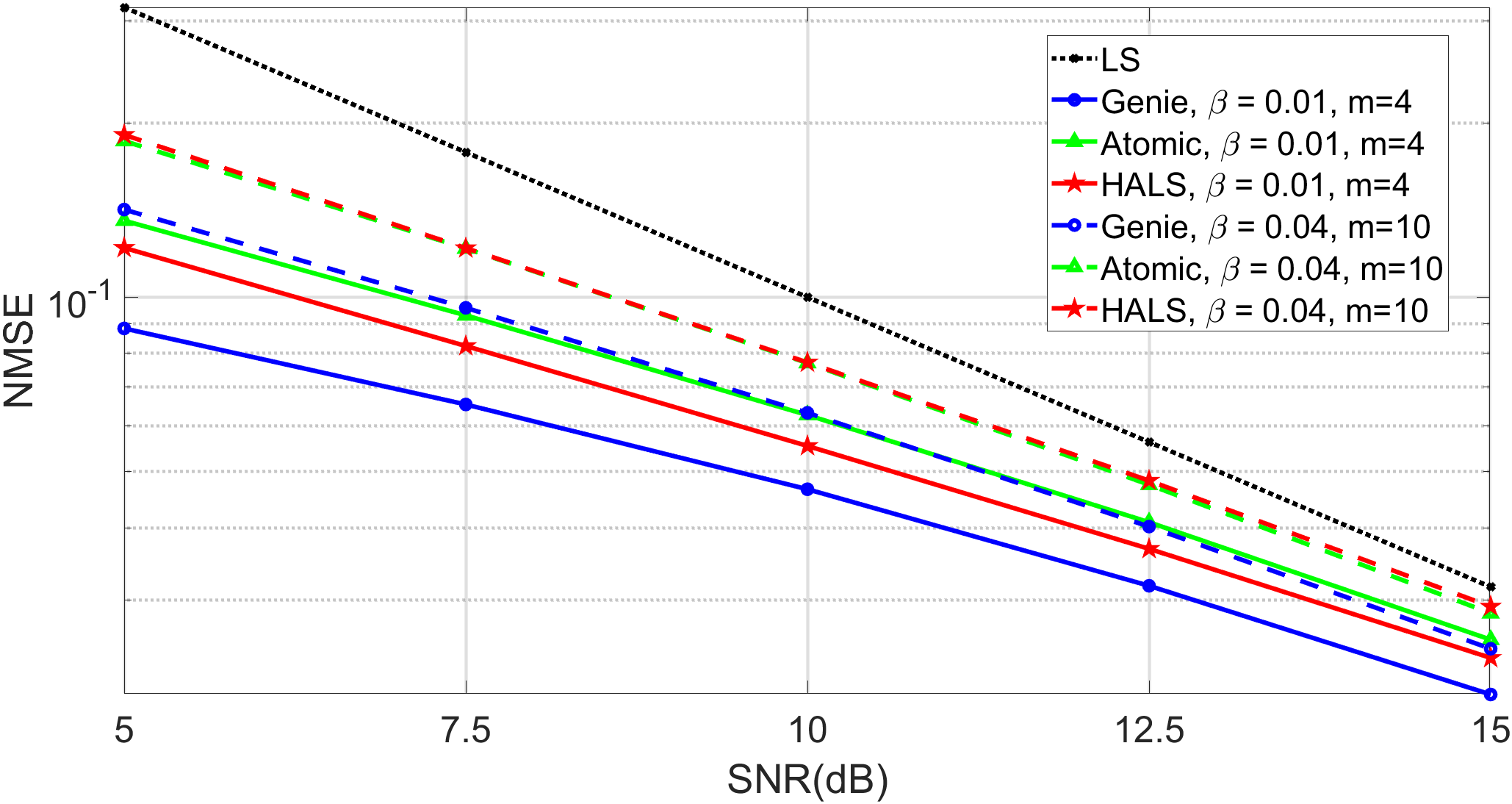}}
  \vspace*{0.1in}
\caption{NMSE versus SNR results with synthetic data. Parameter values are: $N=50$, $L=40$ and $\omega=0.05$ for synthetic channel generation.}
\label{fig:exp1_2}
\end{figure}

We first consider both synthetic and real channel responses in Figure \ref{fig:exp1_1} and \ref{fig:exp1_2} as a function of $\mbox{SNR}$. Fix $N =50$ for all channels. We set $L=40$, $\omega = 0.05$ in the synthetic channels. We consider two different settings for the amplitude of the diffuse components and the sparsity of the specular components $\left(\beta, m\right)=\left\{\left(0.01, 4\right), \left(0.04, 10\right)\right\}$. For the real channel responses, we assume that the number of diffuse components is known and fixed at $\tilde{L} = 200$. Then, we formulate $\boldsymbol{D}$ as $\tilde{\boldsymbol{D}} = \frac{1}{\sqrt{N}}\left[\boldsymbol{a}\left(0\right), \boldsymbol{a}\left( \frac{1}{\tilde{L}}\right), \cdots,\boldsymbol{a}\left(\frac{\tilde{L}-1}{\tilde{L}}\right)\right] \in \mathbb{C}^{N\times \tilde{L}}$. 

As expected,  in Figure~\ref{fig:exp1_1}, the unstructured least-squares channel estimator provides the worst performance for both real and synthetic channels. For the CC2 channel,  the diffuse component energy is relatively large and thus the better model is a purely specular, dense, multipath channel. Thus, HALS offers some improvement over the atomic norm denoiser, but it is not significant. In contrast, the CC13 channel is closer to our proposed model: strong, highly sparse specular paths and modest diffuse scattering, resulting in stronger NMSE performance for HALS. 

These trends are mimicked in the synthetic data experiments seen in Figure~\ref{fig:exp1_2}. When we work on channels with strong diffuse components and dense specular paths, where $\left(\beta, m\right)= \left(0.04, 10\right)$, HALS and ANM are essentially coincident (dashed lines).  However, when we have channels with the HSD model, $\left(\beta, m\right)= \left(0.01, 4\right)$, HALS has measurable improvement over ANM (solid lines). We do see that the critical element to strong performance is knowledge of the support as evidenced by the performance of the genie-aided algorithm (blue curves) which outperforms all other methods on the synthetic data.

\begin{figure}[t!]
  \centerline{\includegraphics[width=8.5cm]{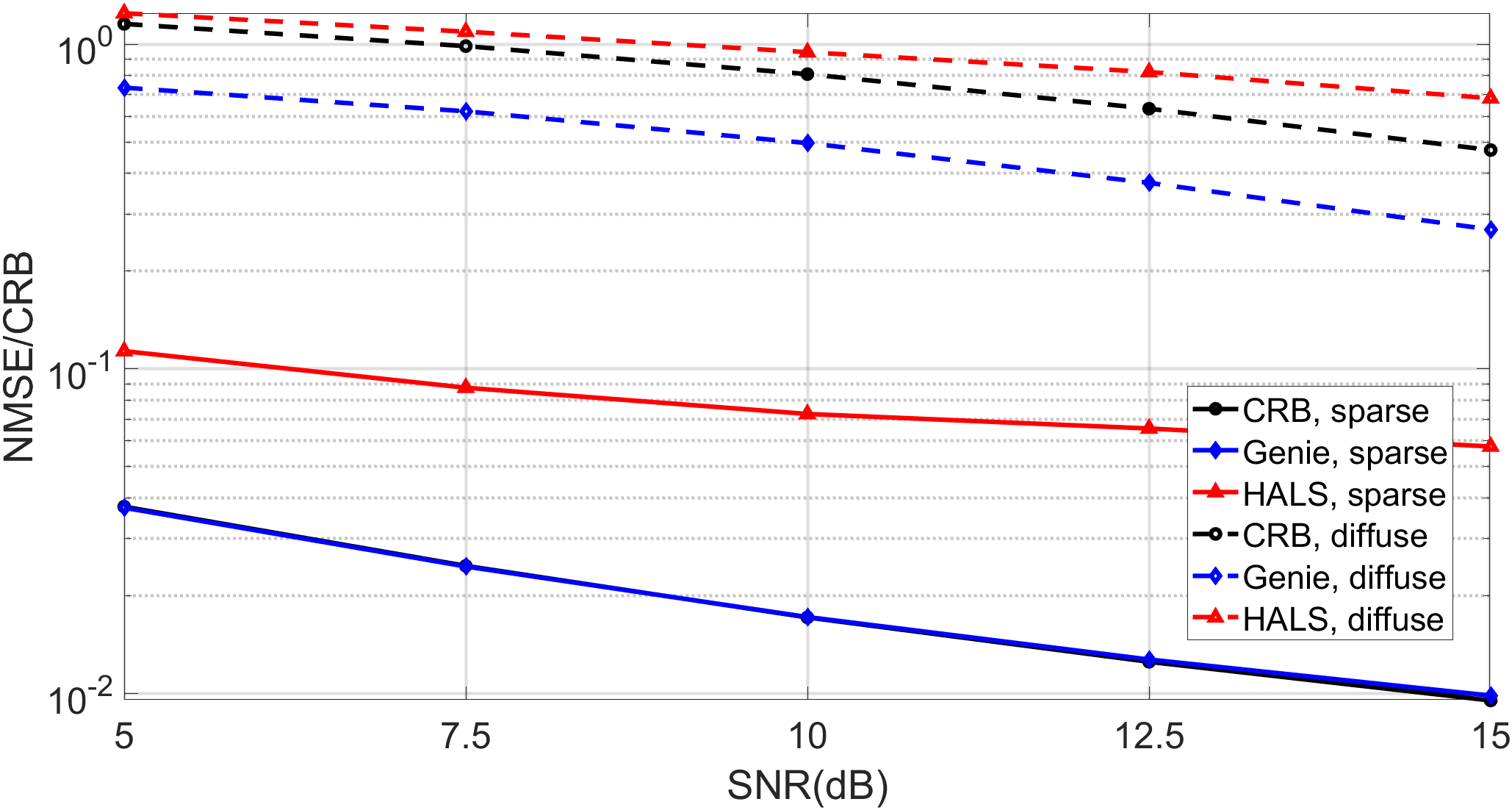}}
    \vspace*{0.1in}
\caption{NMSE/NCRB versus SNR results with HSD-matching synthetic data. Parameter values are: $N=50$, $L=40$, $\omega=0.05$, $\beta = 0.01$ and $m =4$.}
\label{fig:exp2_1}
\end{figure}

We next compare the performance of HALS to the approximate CRBs we have derived in Section~\ref{sec:perform_limit}. We consider performance measures for the sparse and diffuse components separately by examining NMSE and normalized CRB (NCRB) measures for these different components.
In Figure \ref{fig:exp2_1} , we set $\beta = 0.01, m=4$ and this scenario matches our HSD model. The approximate CRB for the diffuse components in \eqref{eq:diffuse_crb} is predictive of the HALS behavior although larger than that of the genie-aided performance.  In contrast, the approximate CRB in \eqref{eq:sparse_crb} for the sparse components is coincident with the genied-aided NMSE and not predictive of the HALS performance.  It appears that the key performance limitation for HALS is knowing the specular component support; the challenge is that diffuse components can be mistaken for sparse components.  We underscore that it is unclear if anything near the genie-aided performance can be achieved for this mixed channel model.

\begin{figure}[t!]
  \centerline{\includegraphics[width=8.5cm]{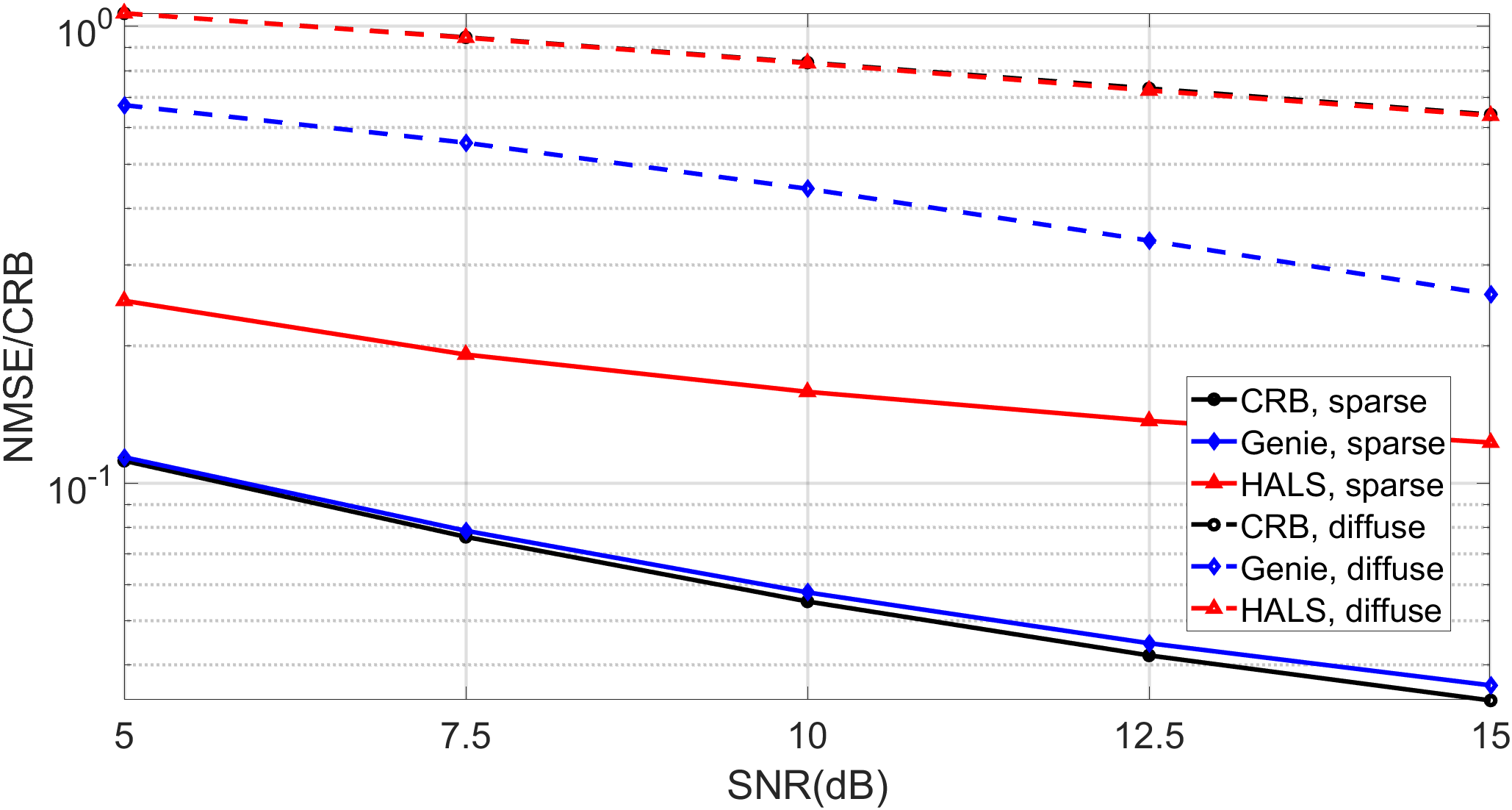}}
    \vspace*{0.1in}
\caption{NMSE/NCRB versus SNR results with a channel model not matched to HSD. Parameter values are: $N=50$, $L=40$, $\omega=0.05$, $\beta = 0.04$ and $m =10$.}
\label{fig:exp2_2}
\end{figure}

In Figure \ref{fig:exp2_2} we set  $\beta = 0.04, m=10$, in which case, the channel resembles a dense multipath channel.  We observe that the trends exhibited in Figure \ref{fig:exp2_1} hold again:  the approximate CRB for the diffuse parts is predictive of the HALS performance and the approximate CRB for the sparse components is coincident with the performance of the genied-aided channel estimator.  The gap between HALS and the genie-aided scheme on the sparse components is relatively larger in this case.

\section{Conclusions}
\label{sec:conclusions}
In this work, we propose the HSD channel framework from continuous time to the frequency domain for OFDM signaling. 
Through a structural analysis, we propose the Hybrid Atomic-Least-Squares (HALS) algorithm by regularizing sparse and diffuse terms with the atomic norm and $l_2$ norm respectively. We further provide a theoretical analysis on the relevant dual problems and exploit properties of primal and dual solutions. These results suggest  a new frequency support estimator. 
Our numerical results show that the proposed HALS algorithm's mean-squared error is upper bounded by the performance of the classical ANM algorithm (ignoring the diffuse components in our HSD model). HALS shows strong improvement over ANM when the underlying channel has a clear hybrid structure between the sparse and diffuse components. Our numerical results on a real dataset agree with the observations from the synthetic data. We have also conducted preliminary analyses to derive approximate CRBs on the NMSE of the sparse and diffuse components separately.  These CRB analyses suggest that it may be possible to further improve the performance of HALS by focusing on an improved support detector for the specular, sparse components.  Additional work is needed both on the channel estimator design and the CRB performance analysis to answer this question.

\bibliographystyle{IEEEbib}
\bibliography{ref}
\end{document}